\documentclass[twocolumn, secnumarabic, amssymb, aps, showpacs, prl]{revtex4-2}

\usepackage{graphics}
\usepackage[dvips]{graphicx}
\usepackage{dcolumn}
\usepackage{xcolor}
\usepackage{amsmath}
\usepackage{caption}
\usepackage{mathrsfs}

\begin{document}

\title{Transition from MOS to Ideal Capacitor Behavior Triggered by Tunneling in the Inversion Population Regime}
\author{Pedro Pereyra}
\affiliation{Departamento de Ciencias B\'asicas, UAM-Azcapotzalco, M\'{e}xico D.F., C. P. 02200, M\'exico}
\date{\today}
\begin{abstract}

An analytical solution to the nonlinear Poisson equation governing the inversion layer in metal–oxide–semiconductor (MOS) structures has recently been obtained, resolving a fundamental challenge in semiconductor theory first identified in 1955. This breakthrough enables the derivation of explicit expressions for relevant physical quantities, such as the inversion-layer width, electric potential, and charge distribution, as functions of gate voltage $V_G$, distance from oxide-semiconductor interface and impurity concentration. These quantities exhibit rapid variation during early-stage inversion but saturate once $V_G$ exceeds the threshold voltage by a few tenths of a volt signaling a transition in the MOS response to gate voltage. The onset of tunneling through the Esaki barrier leads to increased charge accumulation near the interface, reshaping the charge distribution into a two-dimensional profile and shifting the potential drop from the semiconductor to the oxide layer. This reconfiguration resembles the behavior of an ideal parallel-plate capacitor, with charge confined at the interface and the voltage drop localized across the oxide. We analyze this mechanism in detail and demonstrate, through explicit calculations, that the tunneling current through the Esaki-like barrier formed during inversion becomes dominant, effectively superseding classical inversion behavior. These results offer a new analytical foundation for quantum-aware device modeling and inform the design of next-generation MOSFET and tunneling FET architectures.

\end{abstract}
\maketitle

\section{Introduction}

The nonlinear Poisson equation describing the inversion layer in metal-oxide-semiconductor (MOS) structures has remained analytically unsolved for nearly seven decades. This longstanding challenge has prevented a complete theoretical description of inversion-layer behavior, especially under strong inversion conditions. In their foundational 1955 work, Kingston and Neustadter acknowledged the intractability of the problem and proposed focusing on quantities that ``could be evaluated.''\cite{Kingston}. In 1955, Kingston and Neustadter were the first to address the nonlinear Poisson equation for the inversion layer, successfully performing a first integration. Once the electric field $\mathcal{E}(\phi)$ is obtained, the potential $\phi(z)$ can be expressed implicitly as \cite{Tsividis,Pierret}
    \begin{equation}\label{Eq1}
      \int_{\phi(z)}^{\phi_s}\frac{d\hat{\phi}}{\mathcal{E}(\hat{\phi})} = z - z_{\rm surface}.
    \end{equation}
    However, this integral cannot be inverted analytically to yield $\phi(z)$ explicitly and is typically solved by numerical integration. In the decades since, inversion has been modeled indirectly as a two-dimensional electron gas (2DEG), adopting the charge-sheet model,\cite{Brews} and using parallel-plate  capacitor analogies to extract parameters like surface potential, capacitance, and total semiconductor charge.\cite{Temple,Ortiz2005}  While several approximate analytical models have been proposed,\cite{Hauser,Taur2000} and numerical simulations have offered reliable predictions, the absence of exact solutions has limited theoretical insight into the nonlinear regime and its physical consequences.

\begin{figure}
\begin{center}
\includegraphics [width=8.0cm]{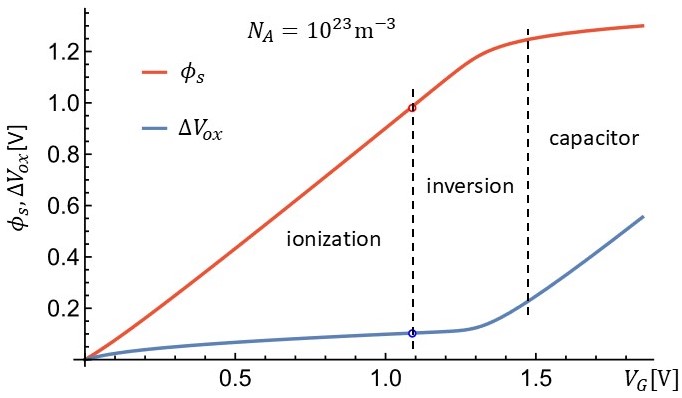}
\caption{The operating regimes of the MOS seen through the behavior of the surface potential $\phi_s$ and the potential drop $\Delta \phi_{ox}$ in the oxide layer, as functions of the gate potential $V_G$.  For this plot, we assume a $p$-type silicon semiconductor and a silicon-oxide layer  2nm thickness.}\label{MOSRegimes}
\end{center}
\end{figure}

A complete analytical solution to this problem has now been obtained.\cite{PereyraSST} The resulting expressions for  the electric potential, inversion-layer width, and charge distribution enable a detailed, parameter-dependent analysis of the inversion process. These quantities are given as explicit functions of gate voltage, position from oxide-semiconductor interface, and material parameters, including oxide thickness and semiconductor impurity (doping) concentration. The later, in particular, plays a role in determining the onset and strength of inversion, and its inclusion allows for predictive modelling across a range of technologically relevant doping profiles. This analytical framework reveals a saturation regime in which inversion-layer characteristics stabilize and the device enters a quantum-dominated phase marked by tunneling through an Esaki-like barrier. This mechanism enhances charge transport from the valence band to the oxide-semiconductor interface, where the charge distribution evolves into a quasi-two-dimensional profile. In this regime, the gate voltage drop increasingly shifts from the semiconductor to the oxide, as illustrated in Fig. \ref{MOSRegimes}.

In the following sections, we revisit the alternative Poisson equation introduced in Ref. \onlinecite{PereyraSST}, whose explicit solution is analyzed here to explore its physical implications. We focus on the transient behavior during the formation of the inversion layer, which begins when the gate voltage reaches the threshold. The process intensifies as the conduction band aligns with the valence band edge, allowing for quantum tunneling through an Esaki-like barrier and promoting charge accumulation. We discuss the saturation observed in the inversion-layer width and electric potential as the gate voltage increases. Together with a numerical evaluation of the tunneling current, this analysis reveals a rapid charge accumulation near the interface and a transition where the potential drop shifts from the semiconductor to the oxide, effectively transforming the MOS structure into a parallel-plate capacitor.

\section{Inversion-Layer Width and Potential Distributions}\label{PEILW}
It was recently shown\cite{PereyraSST} that by taking into account the displacement field and inversion layer currents, the non-linear Poisson equation can be replaced by the alternative differential equation
\begin{equation}\label{NLDEq}
\phi_T \frac{d^2 \phi(z)}{d z^2}=\frac{1}{2}\left(\frac{d \phi(z)}{d z}\right)^2-\frac{1}{2}\xi^2,
\end{equation}
with the same solution as the original Poisson equation. Here $\xi$ is a model dependent parameter, given by
\begin{equation}\label{xi}
\xi^2={\cal E}^2_t+2\frac{\phi_T \rho(\phi_t)}{\epsilon_s}.
\end{equation}
where  $\epsilon_s$ is the semiconductor dielectric constant, $\phi_T=k_BT/e$ the thermal potential, with $k_B$ the Boltzmann constant, ${\cal E}_t$ and $\rho(\phi_t)$ the electric field and the charge density at the threshold $\phi_t=(E_c-E_{Fs})/e \equiv V_{Gu}$, i.e. at the beginning of the inversion layer process, and $E_c$ and $E_{Fs}$ the energy of the conduction band edge and the semiconductor Fermi energy, respectively. The analytical solution of (\ref{NLDEq}), obtained in Ref. \cite{PereyraSST}, is
\begin{equation}\label{ElectPot0}
\phi(z)=\phi_s-2\phi_T\ln\left(\cosh{\frac{\xi (z\!-\!z_0)}{2\phi_T}}\!-\!
\frac{{\cal E}_s}{\xi}\sinh{\frac{\xi (z-z_0)}{2\phi_T}}\right),\hspace{0.3in}
\end{equation}
with $\phi_s$ and ${\cal E}_s$ the electric potential and  electric field at the surface $z=z_0=0$. This method was applied to the Kingstone-Neustadter model, where the charge density is written as\cite{Kingston,Shockley,Tsividis}
\begin{equation}\label{Tsividis2}
\rho(z)\!=\!eN_A\left[e^{-\phi(z)/\phi_T}-1-e^{-2\phi_F/\phi_T}\Bigl(e^{\phi(z)/\phi_T}-1 \Bigr)\right].
\end{equation}
Here $e$ is the electron charge, $N_A$ the impurity concentration,  $\phi_F=E_F/e$ {\color{blue}($\phi_F=E_g/e$)} the intrinsic Fermi potential, and $\phi(z)$ the electric potential. For this density, the model dependent parameter $\xi$ is\cite{Note}
\begin{eqnarray}\label{xi}
\xi^2\!&=&\frac{4eN_A}{\epsilon_s}e^{-\phi_F/\phi_T} \Bigl[2\phi_Te^{-V_{Gu}/2\phi_T}\cosh\frac{V_{Gu} -2\phi_F}{2\phi_T}\cr &+&V_{Gu}\sinh\frac{\phi_F}{\phi_T}\Bigr].
\end{eqnarray}

In modeling charge densities in non-degenerate semiconductors, it is common to approximate the Fermi-Dirac distribution by the Boltzmann distribution. However, when the gate voltage increases beyond the threshold, the conduction and valence bands may align, leading to degeneracy near the surface. In this regime, the Boltzmann approximation becomes invalid. Nonetheless, it is worth noting that Eq.~(2) and its solution (4) remain structurally valid regardless of the specific form of the charge density, though the functional form of the parameter $\xi(\phi)$ would differ. The general reasoning supporting the solution holds even under Fermi-Dirac statistics.

It was shown in Ref. \onlinecite{PereyraSST} that the solution (\ref{ElectPot0}) and the electric field for the Kingstone-Neustadter model, coincide with the numerical solutions of the original non-linear Poisson equation.\cite{Reiter} It was also shown that when applied to the approximate, but solvable, Hauser-Littlejohn model\cite{Hauser}, the corresponding analytical solution (\ref{ElectPot0}) coincides with the Hauser-Littlejon solution.

With the potential distribution (\ref{ElectPot0}) applied to the Kingstone-Neustadter model,  one has not only the electric potential distribution, one also has, by replacing it in (\ref{Tsividis2}), the charge distribution. Before we analyze these  quantities, it is  important to introduce another specific variable that characterizes the inversion layer regime, the inversion-layer width defined as the distance from the oxide-layer interface to the point $z_i$ in which the conduction band edge $E_c-e\phi(z)$ crosses the semiconductor Fermi energy $E_{Fs}$.
\begin{figure}
\begin{center}
\includegraphics [width=8.5cm]{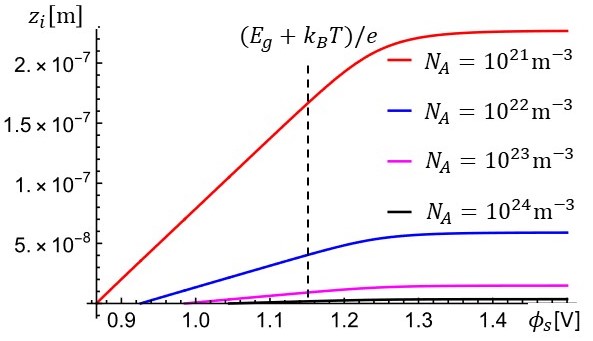}
\caption{The inversion layer width as function of the surface potential $\phi_s$. For these plots we consider $p$-type silicon semiconductors with the indicated doping concentrations and a 2nm-thick silicon-oxide layer.}\label{InvWidth}
\end{center}
\end{figure}
The continuity of the electric potential at $z=z_i$, implies the relation
\begin{equation}\label{DeContzi}
\phi_s-2\phi_T\ln\left(\cosh(az_i/2\phi_T)+
\frac{{\cal E}_s}{a}\sinh(az_i/2\phi_T)\right)=V_{Gu},
\end{equation}
which led to obtain the inversion-layer width
\begin{equation}\label{InversionWidth}
  z_i=\frac{2\phi_T}{\xi}\cosh^{-1}\Bigl( \frac{-e^{\theta/2\phi_T}\xi^2+\sqrt{{\cal E}_s^4-{\cal E}_s^2\xi^2(1-e^{\theta/\phi_T})}}{{\cal E}_s^2-\xi^2}\Bigr).
\end{equation}
Here $\theta=\phi_s-V_{Gu}$. We will start the analysis of the inversion process through the behavior of this quantity.  In figure \ref{InvWidth}, we plot the inversion-layer width $z_i$ as a function of the surface potential $\phi_s$, for different values of the impurity concentration $N_A$. While one can also plot $z_i$  as a function of the gate potential $V_G$, using $\phi_s$ as the independent variable makes the causal relationships more evident, specifically, between the threshold potential, the saturation phenomenon, the alignment with the valence band edge, the onset of tunneling, and the resulting charge accumulation. We will return to this relationship later.

The relation between the gate potential and the surface potential is well defined,  important and given by
\begin{equation}\label{PotDrop}
  V_G=\phi_s+{\cal{E}}_{ox}t_{ox}\hspace{0.2in} {\rm with}\hspace{0.2in} {\cal{E}}_{ox}=-\kappa \frac{d\phi(z)}{dz}\Big|_{z=0}.
\end{equation}
Here $t_{ox}$ is the oxide layer width, ${\cal{E}}_{ox}$ is the electric field in the oxide layer and  $\kappa$ the dielectric ratio. The first term on the right side of (\ref{PotDrop}), i.e. the surface potential $\phi_s$, represents the potential drop in the semiconductor layer and the second term, ${\cal{E}}_{ox}t_{ox}$,  represents the potential drop in the oxide layer. This is generally denoted as  $\Delta V_{ox}$. To be specific and for the specific calculations of the explicit functions just reproduced, we will assume  that the semiconductor is silicon, type $p$\cite{Data}.  In figure \ref{MOSRegimes}, both terms, $\phi_s$ and $\Delta V_{ox}$, are plotted as functions of the gate potential $V_G$.

\section{Saturation Phenomenon in MOS Behavior}

The behavior of the electric potential within the semiconductor, illustrated in figure \ref{MOSRegimes}, where three distinct regimes can be identified, and the limiting growth of the inversion-layer width shown in figure \ref{InvWidth}, both signal the onset of a saturation phenomenon and a transition to a new regime.

At low gate voltages $V_G$, the system resides in the ionization regime, characterized by a surface potential $\phi_s$ below the inversion threshold $\phi_t$. In this regime, the majority of the gate voltage drops across the semiconductor. This drop, quantified by the surface potential $\phi_s$ and corresponding conduction-band bending, increases linearly with $V_G$. The voltage drop across the oxide, $\Delta V_{ox}$, is negligible.

When $\phi_s$ reaches the threshold value $\phi_t$, the inversion layer begins to form, marking the entrance into an intermediate regime, followed by a transition that we will discuss in greater detail below. At higher gate voltages, once the inversion layer is established, the situation reverses: the voltage drop shifts predominantly to the oxide layer. In this inversion regime,  $\Delta V_{ox}$ increases linearly with $V_G$, while the potential drop across the semiconductor becomes minimal.

In the initial part of the inversion regime, when $\phi_s$ lies between $\phi_t$ and approximately 1.25V, the system exhibits linear behavior with respect to $V_G$, as shown in figure \ref{MOSRegimes}. A similar trend is observed in the inversion-layer width plotted in figure \ref{InvWidth} for different doping concentrations. The width increases approximately linearly with $\phi_s$, up to about $\phi_s\sim 1$.2V $\sim (E_g+k_BT)e$

Because different impurity concentrations $N_A$ lead to different threshold potentials, the onset of inversion varies accordingly. Nevertheless, in the limit of large $V_G$, the surface potential $\phi_s$ becomes nearly independent of $N_A$, and the inversion-layer width approaches a maximum. For $N_A \approx 10^{24}$ m$^{-3}$, the saturation width is around 4nm; for $N_A\simeq 10^{22}$m$^{-3}$ it reaches approximately 60nm.

\section{A Transition Triggered by Tunneling}

As previously noted, the saturation phenomenon is also evident in other physical quantities characteristic of the inversion process, whose behavior further confirms the onset of a transition to a new operational regime.

\begin{figure*}
\begin{center}
\includegraphics [width=16.6cm]{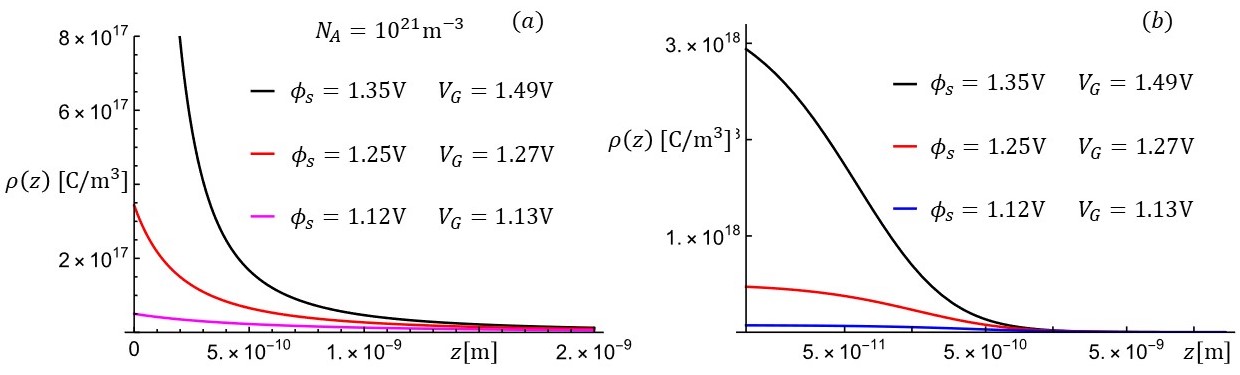}
\caption{The charge concentration $\rho(z)$ as a function of $z$ for three different values of the surface potential. Increasing the gate potential, the charge density varies from almost a constant distribution towards a 2DEG distribution. In panel (b) the same figure but in logarithmic scale on the x-axis.  }\label{ChargeDistr}
\end{center}
\end{figure*}

\begin{figure}
\begin{center}
\includegraphics [width=8.6cm]{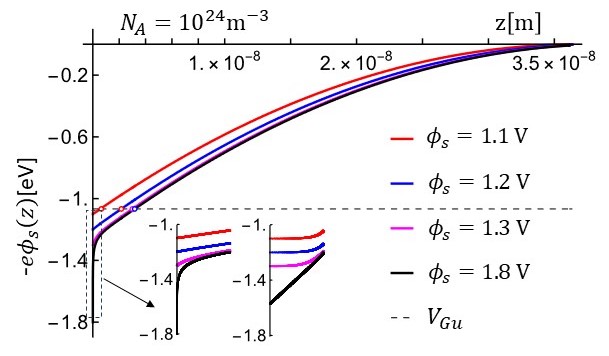}
\caption{Electric potential energy as a function of $z$, for different values of surface potential $\phi_s$, impurity concentration $N_A=10^{24}$/cm$^3 $ and threshold potential $V_{Gu}$. The open circles indicate the inversion-layer width $z_i$. For the surface potentials, whose graphs are plotted here, the corresponding gate potentials are 1.45 V, 1.64 V, 3.05 V and $\sim $ 25700 V, respectively. The inset shows the potential energy in the inversion layer in both normal and logarithmic scales on the x-axis.}\label{PotEnergyfis}
\end{center}
\end{figure}

The physical quantity that led to introduce the inversion process concept is the charge distribution. In figure \ref{ChargeDistr}, we plot $\rho(z)$ as a function of $z$, for $z<z_i$, and for three different values of the surface potential $\phi_s=$1.12, 1.25 and 1.35V. The corresponding values of $V_G$ are also given.  The figure shows that the charge density $\rho(z)/\epsilon_{s}$, which is almost constant for $\phi_s=1$.12V, i.e. when the conduction-band bending reaches the valence band edge, changes rapidly to a distribution that tends to the  two-dimensional electron gas limit. For $\phi_s=$1.35V the density is several orders of magnitude larger and, at the same time, becomes highly localized. This is precisely the range of potentials in which the  width of the inversion layer stops growing. To understand better the features of this transition, let us see now what happens with the electric potential distribution. In figure \ref{PotEnergyfis}, the potential energy $-e\phi (z)$ is plotted as a function of $z$ for  $N_A=10^{24}$m$^{-3}$ and for different values of the surface (or gate) potential. We see in this graph that changing the gate potential from $\sim 3$V to $\sim$ 26000V, has no effect on the shape of the potential energy. This means that even though the accumulation of quasi-free electrons at the semiconductor surface reaches high values, the electric field in the inversion layer remains at its maximum value. This is an important property that will help us to understand, why and how is that the accumulation of charge at the oxide-semiconductor interface, and the shift of the MOS response,  occur.

\begin{figure*}
\begin{center}
\includegraphics [width=15.6cm]{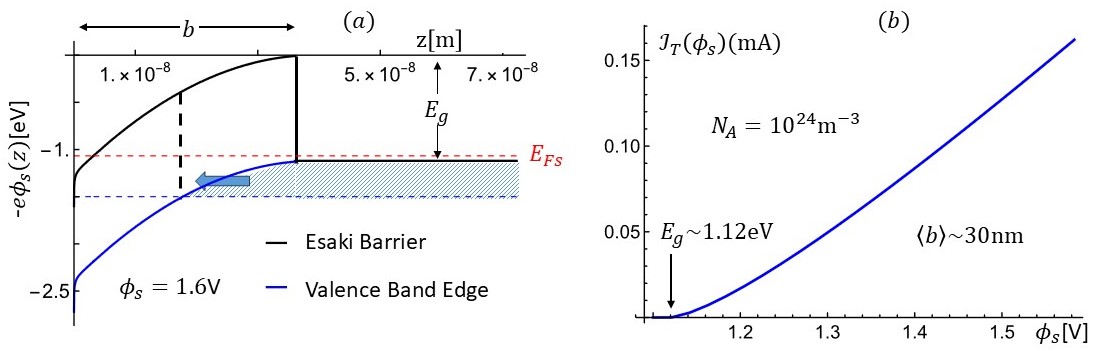}
\caption{Esaki barriers and tunneling current. In (a) electrons of the valence band capable of tunneling through the Esaki barriers whose widths diminish as the gate potential grows. In (b) the tunneling current calculated assuming a triangular shape for the Esaki barrier and an average width of 30nm. }\label{InversionTunneling}
\end{center}
\end{figure*}

As outlined in the abstract and in the introduction, the transition from a growing inversion-layer width to a saturation regime, accompanied by a sharp increase in charge accumulation, begins as soon as quantum tunneling through an Esaki-type barrier becomes accessible to valence band electrons. The abrupt onset of a non-negligible tunneling probability triggers a burst of electron flux toward the narrow, nearly imperceptible potential well that forms between the conduction band edge and the oxide-semiconductor interface (see the inset in Fig. \ref{PotEnergyfis}).

To quantitatively assess this mechanism, namely, the injection of valence-band electrons into the interface region, we performed a simplified calculation of the tunneling current through the Esaki barrier. While a more accurate treatment could employ the WKB approximation to capture the exact barrier profile,\cite{PereyraWeiss} for our purposes, it suffices to approximate the barrier as triangular with an average width $\langle b \rangle =30$nm, corresponding to the impurity concentration used in Fig. \ref{InversionTunneling}(a). Notably, this figure shows that increasing the gate voltage $V_G$ reduces the barrier width, implying that our estimate provides a lower bound for the actual tunneling current. To evaluate the tunneling current, we use the expression $j_{te}(E, V) = e v_i \mathcal{T}(E, V) |\varphi_i|^2 / L^2$, where $\mathcal{T}(E, V)$ is the transmission coefficient at energy $E$ and applied voltage $V$, and $\varphi_i$ is the incident wavefunction. In one dimension, $|\varphi_i|^2 \propto 1/L$, and the density of states is $L / (\pi v_i \hbar)$. These expressions lead to the final form of the current  given by
\begin{equation}\label{tunncurrent}
{\cal I}_T(V)=\frac{e}{\pi \hbar}\int_{-e\phi_s}^{E_v}f(E,T) {\cal T}(E,V)dE
\end{equation}
The tunneling current in figure \ref{InversionTunneling}(b) was obtained after calculating the transmission coefficient as a function of the barrier height, an occupation probability equal to 1 and $E_v$=-1.12eV. This result demonstrates that the tunneling process can effectively supersede the classical inversion mechanism governed by minority carrier concentration. As the gate voltage $V_G$ increases, the effective width of the Esaki barrier becomes narrower, while the electric field on the semiconductor side remains nearly constant. Under these conditions, tunneling enables additional charge accumulation at the interface without further expansion of the inversion-layer width and the electric field within the semiconductor, while the electric field inside the oxide layer increases.
\begin{figure}
\begin{center}
\includegraphics [width=8.4cm]{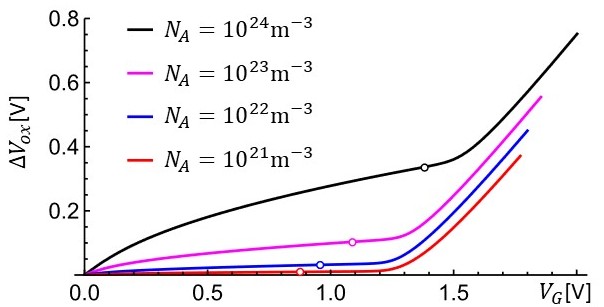}
\caption{The potential drop in the oxide layer as a function of the gate potential. It shows the transition to the capacitor regime of the MOS.  The open circles indicate the beginning of the inversion layer process. }\label{PotentialDropvsVG}
\end{center}
\end{figure}

To visualize this transition, we plot in figure \ref{PotentialDropvsVG} the potential drop as a function of  $V_G$ for different impurity concentrations. The slow growing of $\Delta V_{ox}$, for  $V_G$ smaller than $V_{Gt}$, transforms a few tenths of volt above into the linear behavior of $\Delta V_{ox}$ as a function of $V_G$, a behavior that characterizes the inversion process determined by the inversion tunneling in the capacitor limit of the MOS.

\section{Conclusions}

In this manuscript, we have shown that when the gate voltage exceeds the threshold by a few tenths of a volt, a saturation of the inversion layer width and the electric field within the semiconductor occurs. This is followed by a transition, driven by quantum tunneling through an Esaki-like barrier, leading to rapid charge accumulation near the oxide–semiconductor interface.

This phenomenon was revealed through the analysis of explicit functional forms made possible by a recent theoretical breakthrough: the analytical solution to the nonlinear Poisson equation that governs inversion layers in metal–oxide–semiconductor (MOS) structures—a problem originally posed by Kingston and Neustadter in 1955 and only recently solved in closed form.

The resulting redistribution of charge into a quasi-two-dimensional profile, accompanied by a shift in the voltage drop from the semiconductor bulk to the oxide layer, mirrors the behavior of an ideal parallel-plate capacitor. These findings establish a new theoretical framework for understanding and modeling the interplay between classical inversion and quantum tunneling in advanced MOS devices. This opens new avenues for enhancing gate control, reducing leakage, and improving energy efficiency in future CMOS and tunneling FET architectures.

\acknowledgments

The author is a member of, and gratefully acknowledges financial support from, the Sistema Nacional de Investigadores. The author also thanks Jürgen Reiter for his careful reading and insightful comments.


\begin{thebibliography}{100}
\bibitem{Note} In this paper, for the calculation of $\xi$, we consider the potential at threshold $\phi_t=V_{Gu}$ for $E_{Fs}=E_g-k_BT\ln N_v/N_A$, and $T=300$K. This is slightly more accurate than $\phi_t=2\phi_F$, that was considered in Ref. \cite{PereyraSST}.
\end{thebibliography}
\end{document}